\documentclass[prx,groupedaddress,superscriptaddress,reprint,twocolumn,showpacs,notitlepage]{revtex4-1}

\usepackage{amsfonts}
\usepackage{amsmath}
\usepackage{amssymb}
\usepackage{mathtools}
\usepackage{graphicx}
\usepackage{wrapfig}
\usepackage{color}
\usepackage{mathrsfs}
\usepackage{bbm}
\usepackage{commath}
\usepackage[hidelinks]{hyperref}
\hypersetup{
     colorlinks   = true,
     citecolor    = blue,
     linkcolor    = blue,
     urlcolor     = blue
}
\usepackage{subfigure}
\usepackage{mathptmx}
\usepackage{times}
\usepackage{ulem}

\DeclareMathAlphabet{\pazocal}{OMS}{zplm}{m}{n}

\DeclareMathOperator{\csch}{csch}

\makeatletter
\DeclareFontFamily{OMX}{MnSymbolE}{}
\DeclareSymbolFont{MnLargeSymbols}{OMX}{MnSymbolE}{m}{n}
\SetSymbolFont{MnLargeSymbols}{bold}{OMX}{MnSymbolE}{b}{n}
\DeclareFontShape{OMX}{MnSymbolE}{m}{n}{
    <-6>  MnSymbolE5
   <6-7>  MnSymbolE6
   <7-8>  MnSymbolE7
   <8-9>  MnSymbolE8
   <9-10> MnSymbolE9
  <10-12> MnSymbolE10
  <12->   MnSymbolE12
}{}
\DeclareFontShape{OMX}{MnSymbolE}{b}{n}{
    <-6>  MnSymbolE-Bold5
   <6-7>  MnSymbolE-Bold6
   <7-8>  MnSymbolE-Bold7
   <8-9>  MnSymbolE-Bold8
   <9-10> MnSymbolE-Bold9
  <10-12> MnSymbolE-Bold10
  <12->   MnSymbolE-Bold12
}{}

\newcommand{\ignore}[1]{}
\newcommand{\nobibentry}[1]{{\let\nocite\ignore\bibentry{#1}}}
\newcommand{\re}[1]{\text{Re}\,#1}

\newcommand{\bibfnamefont}[1]{#1}
\newcommand{\bibnamefont}[1]{#1}

\newcommand{\bea}{\begin{eqnarray}}
\newcommand{\eea}{\end{eqnarray}}

\begin{document}

\normalem

\title{Low-temperature thermometry can be enhanced by strong coupling}
\author{Luis A. Correa}
\affiliation{School of Mathematical Sciences and Centre for the Mathematics and Theoretical Physics of Quantum Non-Equilibrium Systems, The University of Nottingham, University Park, Nottingham NG7 2RD, United Kingdom}
\email{luis.correa@nottingham.ac.uk}
\affiliation{Departament de F\'{i}sica, Universitat Aut\`{o}noma de Barcelona, 08193 Bellaterra, Spain}

\author{Mart\'{i} Perarnau-Llobet}
\affiliation{Max-Planck-Institut f\"{u}r Quantenoptik, Hans-Kopfermann-Str. 1, D-85748 Garching, Germany}
\affiliation{Institut de Ci\`encies Fot\`oniques (ICFO), The Barcelona Institute of Science and Technology, 08860 Castelldefels (Barcelona), Spain}

\author{Karen V. Hovhannisyan}
\affiliation{Department of Physics and Astronomy, Ny Munkegade 120, 8000 Aarhus, Denmark}
\affiliation{Institut de Ci\`encies Fot\`oniques (ICFO), The Barcelona Institute of Science and Technology, 08860 Castelldefels (Barcelona), Spain}

\author{Senaida Hern\'andez-Santana}
\affiliation{Institut de Ci\`encies Fot\`oniques (ICFO), The Barcelona Institute of Science and Technology, 08860 Castelldefels (Barcelona), Spain}

\author{Mohammad Mehboudi}
\affiliation{Departament de F\'{i}sica, Universitat Aut\`{o}noma de Barcelona, 08193 Bellaterra, Spain}

\author{Anna Sanpera}
\affiliation{Departament de F\'{i}sica, Universitat Aut\`{o}noma de Barcelona, 08193 Bellaterra, Spain}
\affiliation{Instituci\'o Catalana de Recerca i Estudis Avan\c{c}ats (ICREA), Psg. Llu\'is Companys 23, 08010 Barcelona, Spain} 

\begin{abstract}
We consider the problem of estimating the temperature $ T $ of a very cold equilibrium sample. The temperature estimates are drawn from measurements performed on a quantum Brownian probe \textit{strongly} coupled to it. We model this scenario by resorting to the canonical Caldeira-Leggett Hamiltonian and find analytically the exact stationary state of the probe for arbitrary coupling strength. In general, the probe does not reach thermal equilibrium with the sample, due to their non-perturbative interaction. We argue that this is advantageous for low temperature thermometry, as we show in our model that: (i) The thermometric precision at low $ T $ can be significantly enhanced by strengthening the probe-sampling coupling, (ii) the variance of a suitable quadrature of our Brownian thermometer can yield temperature estimates with nearly minimal statistical uncertainty, and (iii) the spectral density of the probe-sample coupling may be engineered to further improve thermometric performance. These observations may find applications in practical nanoscale thermometry at low temperatures---a regime which is particularly relevant to quantum technologies.
\end{abstract}

\date{\today}
\maketitle

\section{Introduction}

The development of \textit{nanoscale} temperature sensing techniques \cite{carlos2015thermometry} has attracted an increasing interest over the last few years due to their potential applications to micro-electronics \cite{williams1986scanning,aigouy2005scanning,lefevre20053omega}, biochemistry, or even to disease diagnosis \cite{klinkert2009microbial,kucsko2013nanometre,schirhagl2014nitrogen,somogyi2014computational}. In particular, thermometer miniaturization may be taken to the extreme of devising individual \textit{quantum} thermometers \cite{falcioni2011estimate,PhysRevA.88.063609,PhysRevApplied.2.024002,
PhysRevApplied.2.024001,neumann2013high,kucsko2013nanometre,
PhysRevA.93.053619,PhysRevLett.119.090603}.  Using small thermometers, or probes, has the advantage of leaving the sample mostly unperturbed. In contrast, the direct manipulation of the sample, such as e.g. time-of-flight measurements of ultra-cold trapped atoms, is generally destructive and thus, potentially problematic. 

The problem of measuring the temperature $ T $ of an equilibrium sample can thus be naturally tackled  by thermally coupling it to a probe. After equilibration of the probe, one can estimate $ T $ by monitoring some temperature-dependent feature of the probe via a suitable measurement and data analysis scheme. Provided that the heat capacity of the probe is low, one usually assumes that the back-action on the sample can be neglected, and that the probe ends up in a Gibbs state at the sample temperature.  Such a simple picture runs into trouble if the sample is \textit{too cold}, especially when using an individual quantum thermometer: The seemingly natural assumption of the probe reaching equilibrium at the sample temperature might break down at low $ T $. In this limit, the two parties can build up enough correlations to eventually keep the probe from thermalizing
\cite{nieuwenhuizen2002statistical,PhysRevE.86.061132,Ludwig,0034-4885-79-5-056001}. Furthermore, if the probe is \textit{too small}, boundary effects become relevant and need to be taken into account to properly describe equilibration and thermalization \cite{0034-4885-79-5-056001,ferraro2012intensive,PhysRevX.4.031019,1367-2630-17-8-085007}. As a result, thermometry with non-equilibrium quantum probes \textit{inescapably} demands some knowledge about the internal structure of the sample, and the probe-sample coupling scheme.

One could still assume thermalization in the standard sense, owing to a vanishing probe-sample coupling. However, in this limit, the `thermal sensitivity' of the probe, which is proportional to its heat capacity \cite{ll5,benoit1,mandelbrot1989temperature,uffink1999}, drops quickly as the temperature decreases \cite{debye1912theorie}---this is an inherent problem of low-temperature thermometry \cite{de2015local}. The main aim of this article is to show how to fight such a fundamental limitation.
 
Here, we extend quantum thermometry to the \textit{strong coupling} regime, by adopting a fully rigorous description of the probe's dynamics. To that end, we make use of the Caldeira-Leggett Hamiltonian, one of the most paradigmatic dissipation models (see, e.g., \cite{weiss2008quantum}. The equilibrium sample is thus represented by a bosonic reservoir \cite{caldeira1985physica,PhysRevA.31.471} which is \textit{dissipatively} coupled to a single harmonic oscillator, playing the role of the thermometer. We calculate the steady state of the probe exactly and analytically, and show that its low-$ T $ sensitivity is significantly enhanced by increasing the coupling strength. 

In order to quantify the maximum sensitivity attainable by our quantum thermometer, we make use of the quantum Fisher information (QFI) $ \pazocal{F}_T $ \cite{PhysRevLett.72.3439,barndorff2000fisher}. Essentially, $ \pazocal{F}_T $ sets a lower bound on the ``error bars'' $ \delta T \geq 1/\sqrt{M\pazocal{F}_T} $ of any estimate of the temperature of the sample processed from the outcomes of $ M $ independent measurements on the probe \cite{cramer1999mathematical}. Although energy measurements are optimal for thermometry with a probe in thermal equilibrium \cite{PhysRevLett.114.220405}, we will see that these do not harness the potential improvement allowed by strong coupling. 

It is important to stress that we are not limited by any of the simplifying assumptions usually adopted when dealing with open quantum systems, such as the Born-Markov or secular approximations, nor rely on perturbative expansions in the `dissipation strength' \cite{breuer2001time,*breuer2006nonmarkov}. In fact, our methods are totally general and thus, not limited to a specific probe-sample coupling scheme. In particular, we show that our results apply to both Ohmic and super-Ohmic spectra.

We shall now motivate our analysis by illustrating the inherent difficulty of measuring low temperatures in the simplest case. Let us start by considering a quantum probe weakly coupled to a thermal sample, so that its steady state can be written as $\varrho_T=e^{-\beta H_\text{p}}/\mathcal{Z}$, where $H_\text{p}$ is the Hamiltonian of the probe and $ 1/\beta = T $ is the temperature of the sample (in all what follows $ k_B = \hbar = 1 $, and $ \mathcal{Z} $ stands for the partition function). $ \pazocal{F}_T $ can be formally defined as \cite{PhysRevLett.72.3439} 
\begin{equation}
\pazocal{F}_T \coloneqq -2\lim\nolimits_{\delta\rightarrow 0}\partial^2\mathbb{F}(\varrho_T,\varrho_{T+\delta})/\partial\delta^2.
\label{eq:qfi_definition}
\end{equation}
Here, $ \mathbb{F}(\rho,\sigma) \coloneqq \big(\text{tr}\sqrt{\sqrt{\rho}\sigma\sqrt{\rho}}\big)^2 $ stands for the Uhlmann fidelity between states $ \rho $ and $ \sigma $ \cite{uhlmann1976transition,jozsa1994fidelity}, which is a measure of distinguishability between quantum states \cite{fuchs1999cryptographic}. Hence, intuitively, $ \pazocal{F}_T $ gauges the \textit{responsiveness} of the state of the probe to infinitesimal perturbaitons of the global temperature.

The QFI for a single-mode equilibrium thermometer evaluates to 
\begin{align}
\pazocal{F}_T^{(\text{eq})}(\omega) = \tfrac{\omega^2}{4 T^4}\csch^2{\left(\tfrac{\omega}{2 T}\right)},
\end{align}
 which decays exponentially at low $ T $, as can be inferred by expanding it as 
 \begin{align}
  \pazocal{F}_T^{(\text{eq})}(\omega)=\tfrac{\omega^2}{2 T^4} e^{-\omega/ T}+\pazocal{O}(e^{-2\omega/ T} )
  \end{align}
 for $ T/\omega \ll 1$.   This is not specific to harmonic probes, but generally applicable to, e.g., optimized finite-dimensional equilibrium thermometers \cite{PhysRevLett.114.220405}. That is, even an estimate based on the most informative measurements on an optimized \textit{equilibrium} probe has an exponentially vanishing precision as $ T/\omega \rightarrow 0 $. Due to this inherent limitation, devising practical strategies to enhance low-temperature sensitivity becomes ever more relevant, even if these cannot resolve the adverse scaling of $ \pazocal{F}_T $. 

In what follows, we will show that the QFI can improve as the probe-sample coupling increases, and the correlations built up among the two eventually allow the probe to sense a `larger' portion of the sample. First, we obtain the exact analytical (non-equilibrium) steady-state of a harmonic probe as a function of its coupling strength with a sample in thermal equilibrium, to then compute its QFI (see also Ref. \cite{fleming2011exact}). 

\section{The model and its exact solution}\label{sec:model}

Specifically, the Hamiltonian of our probe is just 
\begin{align}
 H_\text{p} = \frac{1}{2}\omega_0^2 x^2 + \frac{1}{2} p^2 
\end{align} (where the mass of the probe is $ m = 1 $), whereas the sample is described as an infinite collection of non-interacting harmonic oscillators
\begin{align}
 H_\text{s} = \sum_{\mu}\frac{1}{2}\omega_\mu^2 m_\mu x_\mu^2 + \frac{1}{2 m_\mu} p_\mu^2. 
 \end{align} 
 The probe-sample coupling is realized by a linear term of the form 
\begin{align}
 H_\text{p--s} = x\sum_{\mu} g_\mu x_\mu .
\end{align}
  In order to compensate exactly for the `distortion' caused on the probe by the coupling to the sample, one should replace $ \omega_0^2 $ with $ \omega_0^2 + \omega_R^2 $ in $ H_\text{p} $ \cite{caldeira1985physica,weiss2008quantum}, where $ \omega_R^2\coloneqq \sum_{\mu}\tfrac{g_\mu^2}{m_\mu\omega_\mu^2} $ (cf. Appendix~\ref{app:SysBath}). 

The coupling strengths between the probe and each of the sample modes are determined by the `spectral density' 
\begin{align}
 J(\omega) \coloneqq \pi \sum_{\mu} \tfrac{g_\mu^2}{2 m_\mu\omega_\mu}\,\delta(\omega-\omega_\mu),
 \end{align}
  which is given a phenomenological analytical form. In the first part of this paper, we shall work with an Ohmic spectral density with Lorentz-Drude cutoff 
  \begin{align}
   J(\omega) = 2 \gamma \omega\,\omega_c^2/(\omega^2+\omega_c^2).
   \end{align} The dissipation strength $ \gamma $ carries the order of magnitude of the couplings $ g_\mu $, and $ \omega_c $ denotes the cutoff frequency, required to ensure convergence.

The following quantum Langevin equation \cite{grabert1984quantum,weiss2008quantum} can be obtained from the Heisenberg equations for $ x $, $ p $, $ x_\mu $ and $ p_\mu $:
\begin{equation}
\ddot{x}(t) + (\omega_0^2+\omega_R^2) x(t) - x(t) \ast \chi(t) = F(t).
\label{eq:qle}
\end{equation}
The first two terms in the left-hand side of Eq.~\eqref{eq:qle} correspond to the coherent dynamics of a free harmonic oscillator of squared frequency $ \omega_0^2+\omega_R^2 $ (the dots denote time derivative), while the incoherent superposition of all environmental modes, encompassed in $ F(t) $, plays the role of a driving force with $ \langle F(t) \rangle = 0 $ (see Appendix~\ref{app:HtoL}). The convolution 
\begin{align} x(t) \ast \chi(t) \coloneqq \int_{-\infty}^\infty \dif s\,\chi(t-s) x(s) 
\end{align} brings memory effects into the dissipative dynamics. Here, 
\begin{align} \chi(t) \coloneqq \frac{2}{\pi}\Theta(t)\int_0^\infty\dif\omega\,J(\omega)\sin{\omega t},
\end{align}
 where $ \Theta(t) $ stands for the step function. 

It is important to remark that Eq.~\eqref{eq:qle} is \textit{exact}. The only assumption that we make when solving it is that probe and sample start uncorrelated at $ t_0\rightarrow-\infty $, i.e. in $ \varrho\otimes\sigma_T $, where $ \sigma_T $ is the Gibbs state of the sample at temperature $ T $. The initial state of the probe $ \varrho $ is arbitrary. However, since the Hamiltonian $ H $ is overall quadratic in positions and momenta, its stationary state is Gaussian, and thus, completely determined by its first and second-order moments: $ \langle R_i(t) \rangle $ and 
\begin{align}
\sigma_{ij}(t',t'') \coloneqq \frac12\langle\lbrace R_i(t'), R_j(t'') \rbrace\rangle  ,
\end{align} where $ R = (x,p) $ \cite{grabert1984quantum}. The notation $ \langle\cdots\rangle $ stands here for average on the initial state and $ \lbrace\cdot , \cdot\rbrace $ denotes anti-commutator. Since $ \langle F(t) \rangle = 0 $, the stationary first-order moments vanish (see Appendix~\ref{app:solvingQLE}).

One may now take the Fourier transform ($ \tilde{f}(\omega) \coloneqq \int_{-\infty}^\infty \dif t\, f(t) e^{i\omega t} $) in Eq.~\eqref{eq:qle}, and solve for $ \tilde{x}(\omega) $, which yields 
\begin{align}
 \tilde{x}(\omega) = \frac{\alpha(\omega)}{\tilde{F}(\omega)},
\end{align} where $ \alpha(\omega) \coloneqq \omega_0^2+\omega_R^2-\omega^2-\tilde{\chi}(\omega) $. The position correlator $ \sigma_{11}(t',t'') $ can be thus cast as
\begin{equation}
\sigma_{11}
 =\iint_{-\infty}^{\infty}\frac{\dif\omega'\dif\omega''}{8\pi^2}\,e^{-i(\omega' t'+\omega''t'')}\,\langle \{\tilde{x}(\omega'),\tilde{x}(\omega'')\} \rangle,
\label{eq:covariance_xx}
\end{equation}
whereas $ \sigma_{22} $ may be calculated similarly by noticing that $ \langle\lbrace \tilde{p}(\omega')\tilde{p}(\omega'') \rbrace\rangle = -\omega'\omega''\langle\lbrace\tilde{x}(\omega')\tilde{x}(\omega'')\rbrace\rangle $. The remaining covariances are $ \sigma_{12} = \sigma_{21} = 0 $ (cf. Appendices \ref{app:solvingQLE} and \ref{app:covariance}).

Hence, all we need to know is the power spectrum of the noise $ \langle\lbrace \tilde{F}(\omega')\tilde{F}(\omega'') \rbrace\rangle $ and the Fourier transform of the dissipation kernel $ \tilde{\chi}(\omega) $. Since the sample was prepared in a Gibbs state, one can show that the noise is connected to the dissipation kernel through the following fluctuation-dissipation relation (cf. Appendix~\ref{app:FDR}) 
\begin{equation}
\langle\{\tilde{F}(\omega'),\tilde{F}(\omega'')\}\rangle = 4\pi\,\delta(\omega'+\omega'') \coth{(\tfrac{\omega'}{2T})}\,\text{Im}\,\tilde{\chi}(\omega').
\label{eq:fluct_dis}
\end{equation}
For our spectral density, we obtain $ \tilde{\chi}(\omega) = 2\gamma\omega_c^2/(\omega_c-i\omega) $ (cf. Appendix~\ref{app:FDR}).

\begin{figure}
	\includegraphics[width=0.95\linewidth]{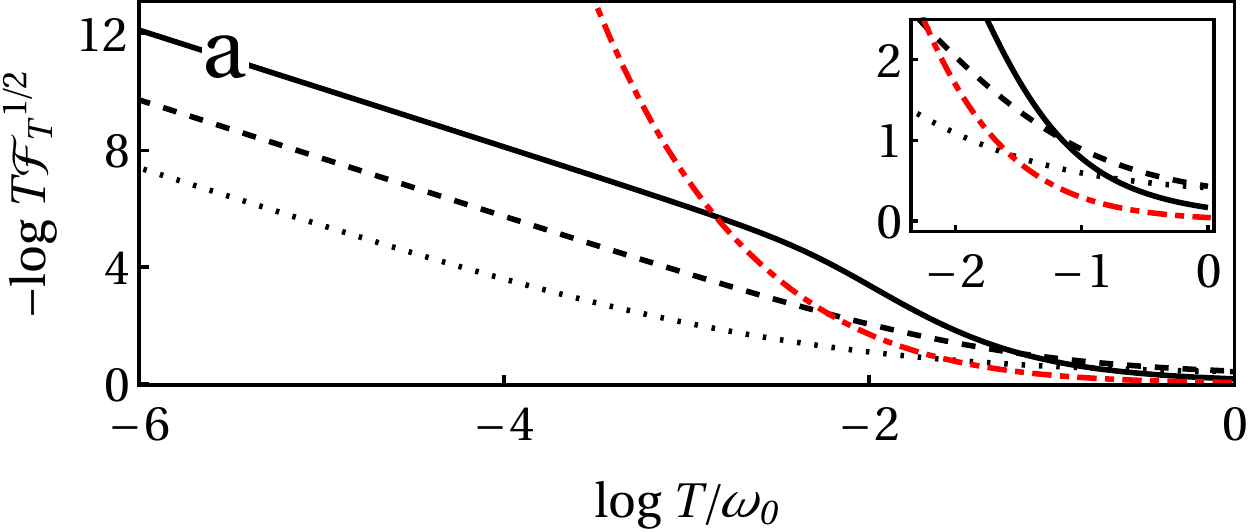}
	\includegraphics[width=0.95\linewidth]{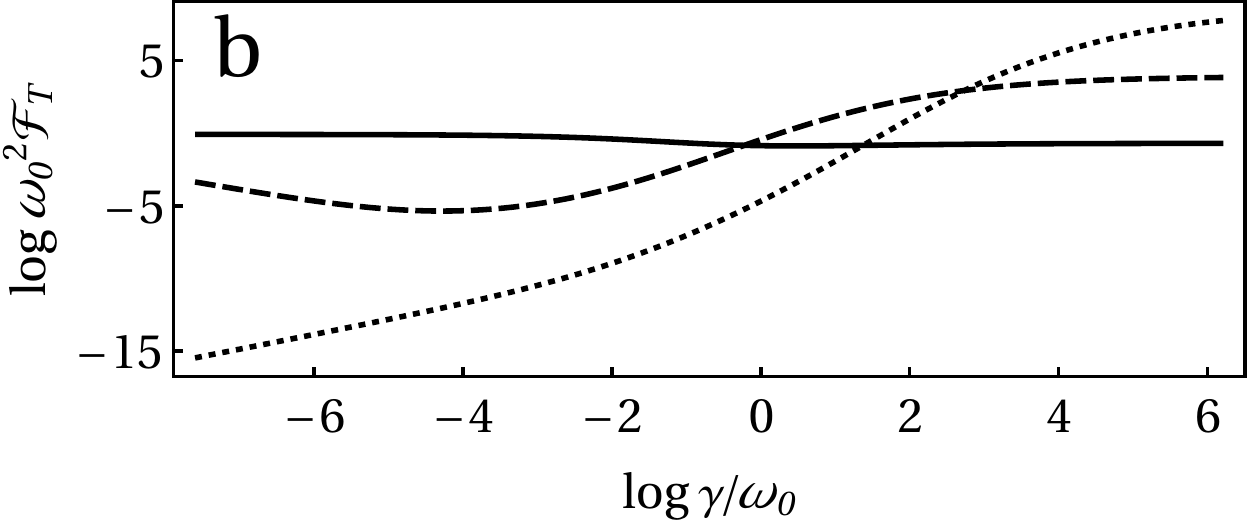}
	\caption{(color online) (a) Log-log plot of the best-case relative error $ \delta T/T = 1/(T\sqrt{\pazocal{F}_T}) $ vs. the sample temperature $ T $ for different dissipation strengths $ \gamma $; namely, $ \gamma/\omega_0 = 0.1 $ (solid black), $ \gamma/\omega_0 = 1 $ (dashed black), and $ \gamma/\omega_0 = 5 $ (dotted black). The relative error of a single-mode probe at thermal equilibrium (dot-dashed red) has been super-imposed for comparison. $ \delta T/T $ diverges as $ T\rightarrow 0 $; while for the thermal mode it would diverge exponentially, our exact solution yields $ \delta T/T \sim T^{-2} $ at low $ T $. Whenever $ T/\omega_0 \ll 1 $, increasing the dissipation strength results in a significant reduction of the minimum $ \delta T/T $. On the contrary, at larger temperatures, the best-case relative error need not be monotonically decreasing with $ \gamma $. This is shown in the inset, which zooms into the bottom-right corner of the plot. (b) Log-log plot of $ \pazocal{F}_T $ as a function of $ \gamma $ for $ T = 1 $ (solid), $ T = 0.1 $ (dashed), and $ T = 0.01 $ (dotted). It becomes again clear that, while not strictly monotonic in $ \gamma $, the QFI \textit{always} grows with the dissipation strength for $ \gamma/\omega_0 \gtrsim 1 $ at $ T/\omega_0 \ll 1 $. Furthermore, as $ T/\omega_0\rightarrow 0 $, we observe such a sensitivity enhancement at arbitrarily weak probe-sample coupling. In both cases $ \omega_c = 100\,\omega_0 $ and $ \hbar = k_B = \omega_0 = 1 $} 
	\label{fig1}
\end{figure}

Putting together the pieces from the above paragraphs, we can compute the steady-state covariances $ \sigma_{ij}(t,t) $ \cite{PhysRevA.86.012110,PhysRevA.88.012309,*PhysRevA.88.042303,*PhysRevE.91.062123} (recall that $ t_0 \rightarrow -\infty $). Importantly, our choice of $ J(\omega) $ makes it possible to evaluate the covariances analytically (see Appendix~\ref{app:explicit-ohmic}). These may be collected into the $ 2\times 2 $ matrix $ \boldsymbol{\sigma} $, which provides a full description of the (Gaussian) non-equilibrium asymptotic state \cite{0503237v1}.

\section{Enhanced thermometry at low $ T $}\label{sec:thermometry}

\subsection{Dissipation-driven thermometric enhancement}\label{sec:dissipation}

We can now calculate $ \pazocal{F}_T $ from Eq.~\eqref{eq:qfi_definition}, using the fact that the Uhlmann fidelity between two single-mode Gaussian states with covariance matrices $ \boldsymbol{\sigma}_1 $ and $ \boldsymbol{\sigma}_2 $ is given by 
\begin{align} \mathbb{F}(\boldsymbol{\sigma}_1,\boldsymbol{\sigma}_2) = 2\,(\sqrt{\Delta+\Lambda}-\sqrt{\Lambda})^{-1},
\end{align}
where $ \Delta\coloneqq 4\det{(\boldsymbol{\sigma}_1+\boldsymbol{\sigma}_2)} $ and $ \Lambda \coloneqq (4\det{\boldsymbol{\sigma}_1}-1)(4\det{\boldsymbol{\sigma}_2}-1) $ \cite{0305-4470-31-15-025}. In Fig.~\ref{fig1}(a) we plot the the best-case relative error $ \delta T/T = 1/(T\sqrt{\pazocal{F}_T}) $ (disregarding the factor $ 1/\sqrt{M} $) versus the temperature of the sample, for different dissipation strengths $ \gamma $. We see how, at low $ T $, the performance of our thermometer is significantly improved by strengthening its coupling to the sample. However, the QFI does not increase monotonically with $ \gamma $, as illustrated in Fig.~\ref{fig1}(b). Instead, only at cold enough $ T $ is the performance of the probe monotonically enhanced by sufficiently strengthening the probe-sample interaction. In the limiting case of approaching zero temperature, such dissipation-assisted enhancement can be attained at arbitrarily low probe-sample coupling.  

It is necessary to specify what we mean by `cold enough' and `sufficiently strong' in this context. The central energy scale of our problem is set by the frequency of the probe $ \omega_0 $. We say that the sample is `cold' whenever $ T/\omega_0 \ll 1 $ so that the probe has a very low thermal population. On the other hand, we say that the coupling is `strong' whenever it is non-perturbative; that is, when $ \gamma/\omega_0 \gtrsim 1 $. In this situation, the probe will certainly end up in a non-equilibrium steady state \cite{grabert1984quantum}. Thus going back to Fig.~\ref{fig1}(b), we see that, provided that $ T/\omega_0 \ll 1 $ and $ \gamma/\omega_0 \gtrsim 1 $, $ \pazocal{F}_T $ increases \textit{monotonically} with the dissipation strength. Hence, the probe-sample coupling can be thought of as a relevant control parameter in practical low-temperature quantum thermometry. This is our main result.

It is worth stressing that even though, in the above, we have resorted to an Ohmic spectral density with algebraic high-frequency cutoff, the exact same qualitative behaviour follows from a spectral density with exponential cutoff $ J_s(\omega) \coloneqq \frac{\pi}{2}\gamma \omega^s\omega_c^{1-s} e^{-\omega/\omega_c} $ and a tunable `Ohmicity' parameter $ s $. In particular, in Appendix~\ref{app:kernel}, we give full details on how to solve the ubiquitous super-Ohmic case $ s > 1 $.  

\begin{figure}[b!]
\includegraphics[width=0.47\linewidth]{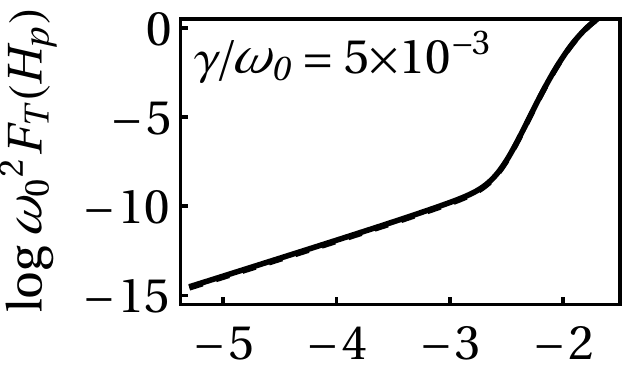}\includegraphics[width=0.47\linewidth]{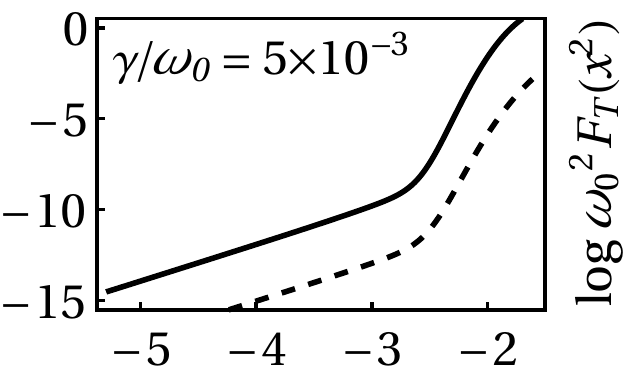}\\
\includegraphics[width=0.47\linewidth]{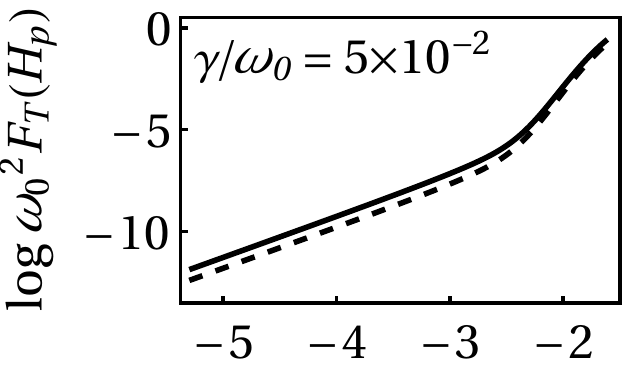}\includegraphics[width=0.47\linewidth]{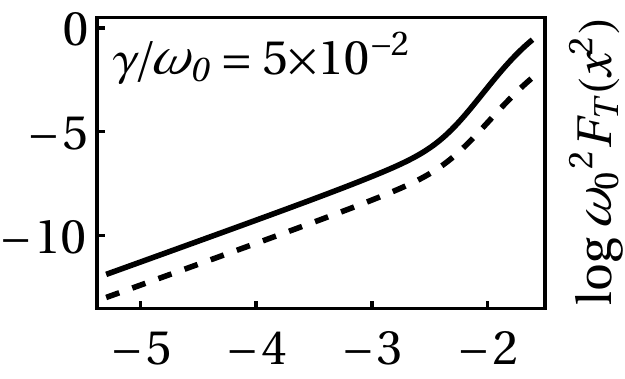}\\
\includegraphics[width=0.47\linewidth]{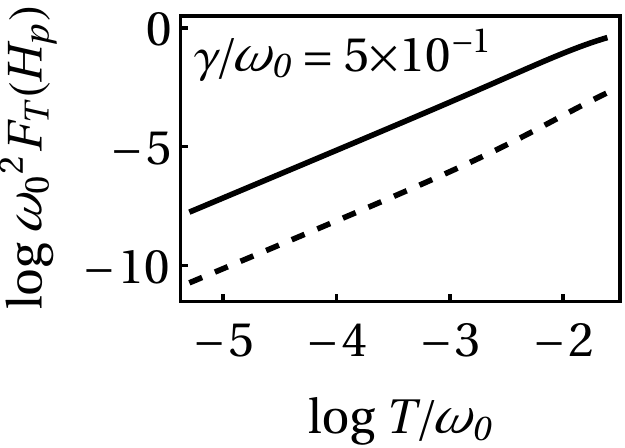}\includegraphics[width=0.47\linewidth]{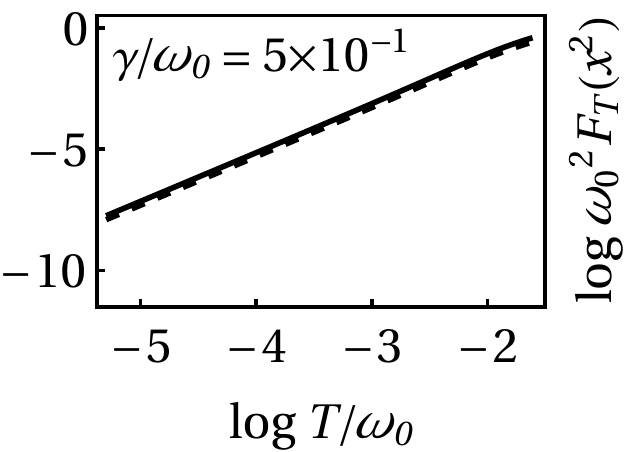}
\caption{Log-log plot of the QFI $ \pazocal{F}_T $ (solid black on all panels), thermal sensitivity of the energy of the probe $ F_T(H_\text{p}) $ (dashed black on the left-hand panels), and $ F_T(x^2) $ (dashed black on the right-hand panels), for different values of the dissipation strength: $ \gamma/\omega_0 = 5\times 10^{-3}$ (top), $\gamma/\omega_0 = 5\times 10^{-2}$ (middle), and $\gamma/\omega_0 = 0.5 $ (bottom). Note that the thermal sensitivity $ H_\text{p} $ is deterred as the dissipation strength grows, whilst $ x^2 $ becomes a quasi-optimal temperature estimator. As in Fig.~\ref{fig1}, $ \omega_c = 100 $ and $ \hbar=k_B=\omega_0=1 $.} 
\label{fig2}
\end{figure}

\subsection{How to exploit strong dissipation in practice}\label{sec:exploit}

Thus far, we have shown how strong coupling may improve the ultimate bounds on thermometric precision at low temperatures. However, we have not yet discussed how to saturate those bounds in practice. We therefore need to find observables capable of producing temperature estimates that approach closely the precision bound set by the QFI.

In general, a temperature estimate based on $ M $ independent measurements of some observable $ O $ on the steady state of the probe has uncertainty $ \delta T \geq 1 / \sqrt{M\,\mathcal{F}_T(O)} $, where $ {\mathcal{F}}_T(O) $ stands for the `classical Fisher information' of $ O $ \cite{barndorff2000fisher}. This may be lower-bounded by the `thermal sensitivity' 
\begin{align} F_T(O) \coloneqq  \frac{\vert\partial_T\langle O \rangle \vert^2}{(\Delta O)^2} \leq \mathcal{F}_T(O) \leq \pazocal{F}_T \equiv \sup_O\,F_T(O) 
\end{align} \cite{PhysRevLett.72.3439,toth2014quantum}. Here, $ \Delta O \coloneqq \sqrt{\langle O^2 \rangle-\langle O \rangle^2} $ denotes standard deviation on the stationary state of the probe. The observable for which $ F_T(O) $ is maximized (i.e. $ F_T(O) = \mathcal{F}_T(O) = \pazocal{F}_T $) commutes with the so-called `symmetric logarithmic derivative' (SLD) $ L $, which satisfies $ \partial_T\varrho = \frac12(L\varrho+\varrho L) $. For instance, in the case of an equilibrium probe, i.e. $ \varrho_T \propto \exp{(-H_\text{p}/T)} $, one has $ [L,H_\text{p}] = 0 $. Consequently, a complete projective measurement on the energy basis renders the best temperature estimate. However, as shown in Fig.~\ref{fig2}, when the strength of the interaction with the sample increases, energy measurements become less and less informative about the temperature of the sample---the larger the dissipation strength $ \gamma $, the smaller $ F_T(H_\text{p})/\pazocal{F}_T $. Estimates based on energy measurements seem thus incapable of exploiting the extra low-temperature sensitivity enabled by the strong dissipation. 

In searching for a more suitable measurement scheme, one can look at the SLD: Since $ \varrho_T $ is an undisplaced Gaussian, $ L $ will be a quadratic form of $ x^2 $ and $ p^2 $ \cite{monras2013phase}. Due to our choice for the probe-sample coupling ($ x\sum_{\mu}g_\mu x_\mu $), the steady state $ \varrho_T $ becomes squeezed in the position quadrature at $ T/\omega_0 \ll 1 $ and $ \gamma/\omega_0 \gtrsim 1 $ \cite{grabert1984quantum,lampo2017bose}. Interestingly, we observe that $ \langle x^2 \rangle $ is much more sensitive to temperature changes in this regime than $ \langle p^2\rangle $. We thus take $ O = x^2 $ as an ansatz for a quasi-optimal temperature estimator. $ F_T(x^2) $ is also plotted in Fig.~\ref{fig2}, where we can see how it does approach closely the ultimate bound $ \pazocal{F}_T $ as $ \gamma $ grows (at $ T/\omega_0 \ll 1 $). This numerical observation can be confirmed by taking the low-temperature limit on the analytic stationary covariances (see Appendix~\ref{app:explicit-ohmic-limit}). 

Measuring the variance of the most relevant quadrature of a Brownian thermometer is therefore a practical means to exploit the thermometric advantage provided by strong dissipation at low temperatures. Putting forward an experimental proposal to demonstrate this dissipation-driven improvement is, however, beyond the scope of this paper. It is worth mentioning that quadratures of trapped particles are either directly measurable \cite{bastin2006measure} or accessible via state tomography \cite{PhysRevA.54.R25,PhysRevA.53.R1966}, and that systems such as an impurity in a BEC may admit a Caldeira-Leggett description \cite{lampo2017bose}. 

\begin{figure}
\includegraphics[width=0.95\linewidth]{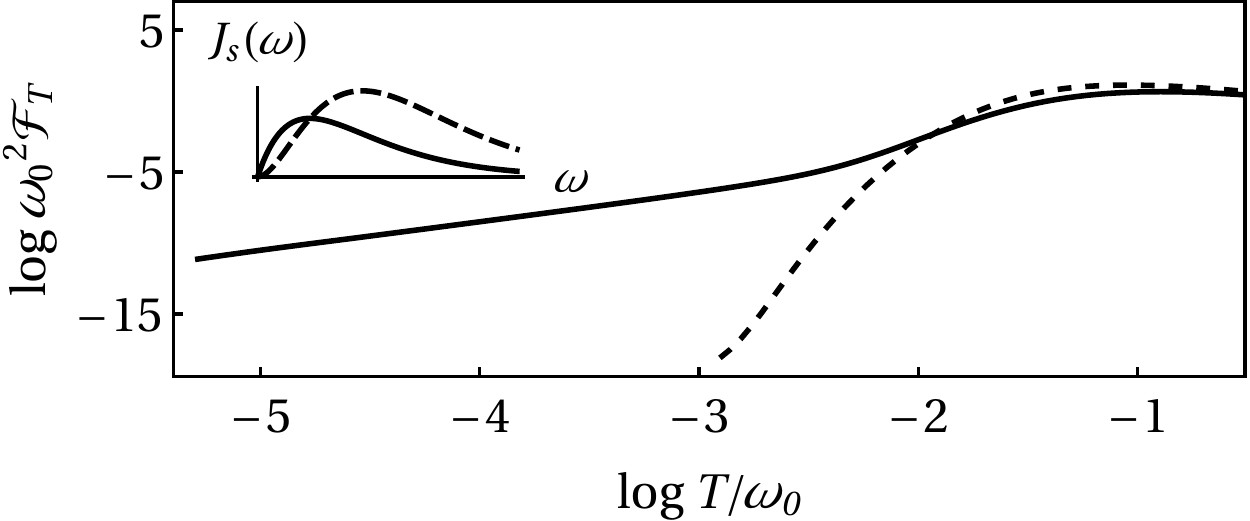}
\caption{Log-log plot of the QFI $ \pazocal{F}_T $ as a function of temperature for Ohmic (solid) and super-Ohmic (dashed) spectral density $ J_s(\omega) $ with exponential high-frequency cutoff ($ s = 1 $ and $ s = 2 $, respectively). In the inset, both spectral densities are compared. Note that the Ohmic form largely outperforms the super-Ohmic one at low temperatures ($ \gamma/\omega_0 = 0.1 $, $ \omega_c = 100\omega_0 $, and $ \hbar = k_B = \omega_0 = 1 $).} 
\label{fig3}
\end{figure}

\section{Discussion and conclusions}\label{sec:discussion}

We shall now give an intuition about the origin of the observed dissipation-driven enhancement. To that end, let us consider not just the marginal of the probe but the global state of probe \textit{and} sample. For simplicity we can model them as a finite $ N $-mode `star system', comprised of a central harmonic oscillator (playing the role of the probe), linearly coupled to $ N-1 $ independent peripheral oscillators with arbitrary frequencies (representing the sample). Let us further prepare the $ N $-mode composite in a Gibbs state at the sample temperature $ T $. Indeed, when such system is at thermal equilibrium, and provided that the number of modes $ N $ is large enough, the marginal of the central oscillator approximates well the \textit{actual} steady state of the probe \cite{PhysRevE.86.061132} (cf. Appendix~\ref{app:freqsvscoupling}). 

As we show in Appendix~\ref{app:freqsvscoupling}, the frequencies of the lowermost normal modes of the global star system always \textit{decrease monotonically} as the overall magnitude of the coupling strengths increases. If the temperature $ T $ was so low that not even the first harmonic could get thermally populated, the sensitivity of the entire system and, by extension, also that of the central probe, would vanish. However, one could populate the first few normal modes by strengthening the couplings, as their frequencies would then decrease [cf. Fig.~\ref{fig1}(b)]. It is this effect which ultimately enables temperature sensing at low $ T $. The magnitude of the enhancement is dictated by the specific frequency distribution of the probe-sample couplings which, in turn, determines the spectrum of the normal modes of the global system.

From the above reasoning it follows that the \textit{shape} of the spectral density $ J(\omega) $ could, in principle, be tailored to render more precise low-temperature probes. To see that this is indeed the case, we shall adopt a generic spectral density of the form $ J_s(\omega) \coloneqq \frac{\pi}{2}\gamma \omega^s\omega_c^{1-s} e^{-\omega/\omega_c} $. We can thus compare the performance of a single-mode thermometer coupled to the sample through an Ohmic ($ s=1 $) and a super-Ohmic ($ s > 1 $) spectral density. Importantly, the dissipation kernel $ \tilde{\chi}(\omega) $ needs to be re-calculated due to the change in spectral density \cite{PhysRevE.91.062123} (see Appendix~\ref{app:kernel}). Note as well that now $ \omega_R^2 = \gamma\omega_c\Gamma(s)$, where $ \Gamma(z)\coloneqq\int_0^\infty \dif t\, t^{z-1} e^{-t} $ is Euler's Gamma function. In Fig.~\ref{fig3} we can see how the Ohmic spectral density offers a clear advantage over the super-Ohmic one at low temperatures. This is in line with our qualitative argument explaining the dissipation-driven enhancement in precision: A thermometer coupled more strongly to the lower frequency modes of the sample (i.e. the only ones substantially populated at low $ T $) should perform better.

As a final remark, we note that, since the equilibrium state of the probe corresponds to the marginal of a global thermal state \cite{PhysRevE.86.061132}, we can think of our results as an instance of thermometry on a macroscopic sample through local measurements, as studied in \cite{de2015local}. While local thermometry in translationally-invariant \textit{gapped} systems is exponentially inefficient at low temperatures \cite{hovhannisyan2017low}, our exact results display a polynomial decay $ \pazocal{F}_T \sim T^{-2} $ as $ T\rightarrow 0 $. Such an exponential advantage can be related to the fact that the Caldeira-Leggett model maps into a \textit{gapless} harmonic chain \cite{hovhannisyan2017low}. This polynomial behaviour holds for both Ohmic and super-Ohmic spectra. 

The aim of this study has been threefold: (i) We have shown that the thermal sensitivity of a single-mode bosonic probe can be boosted by increasing the strength of its dissipative coupling to the sample under study, (ii) we have provided a concrete and feasible measurement scheme capable of producing nearly optimal temperature estimates in the relevant regime and, finally, (iii) we have suggested that the spectral density of the probe-sample coupling can be set to play an active role in enhanced low-temperature quantum thermometry. It is worth emphasizing that all our results are exact, irrespective of the relative ordering of the various time scales involved in the problem. In particular, observation (iii) calls for a more in-depth analysis of the potential role of reservoir engineering techniques \cite{kofman1994spontaneous,PhysRevLett.77.4728} or even dynamical control \cite{PhysRevApplied.5.014007} in enhanced low-$ T $ quantum thermometry and will be the subject of further investigation. 
\newline

\noindent \emph{Acknowledgements.--} We thank A. Ac\'{i}n, A. A. Valido, E. Bagan, A. Monras, and D. Alonso for fruitful discussions. We acknowledge funding from the European Research Council (ERC) Starting Grant GQCOP (Grant No. 637352), the Spanish MINECO (Projects FIS2013-40627-P, FIS2016-80681-P, FOQUS FIS2013-46768-P, QIBEQI FIS2016-80773-P, and Severo Ochoa SEV-2015-0522), the Generalitat de Catalunya (CIRIT Projects 2014 SGR 966, SGR 875, and CERCA Program), CELLEX-ICFO-MPQ Research Fellowships, the Villum Fonden, the ``la Caixa''-Severo Ochoa Program, and the COST Action MP1209.

\appendix

\section{Some remarks on the probe-sample coupling}
\label{app:SysBath}

Let us start by briefly commenting on the renormalization of the frequency of the probe. Splitting the Hamiltonian into a potential and a kinetic term $ H = U(x,x_\mu) + K(p,p_\mu)$, one can see that effective potential \textit{felt} by the probe is given by $ U(x,x^\star_\mu) $, where $ x^\star_\mu = -\tfrac{g_\mu x}{m_\mu\omega_\mu^2}$ [i.e. $ \partial_{x_\mu} U = 0 $ at $ x_\mu^\star $]. This is $ U(x,x^\star_\mu) = \frac{1}{2}(\omega_0^2-\omega_R^2)x^2 $. As a result, the high temperature limit of the reduced steady state of the probe obtained from the \textit{bare model} $ H = H_\text{p} + H_\text{s} + H_\text{p--s} $ is $\text{tr}_\text{s}\,\rho \propto \exp{\big(-\frac{1}{2 T}(\omega^2-\omega_R^2)x^2-\frac{1}{2 T} p^2\big)} $, which may differ significantly from the corresponding thermal state $ \varrho_T = Z^{-1}\exp{(-H_\text{p}/T)} $ if the couplings $ g_\mu $ are strong. To correct this, one must introduce the frequency shift $ \omega_R^2 $ in $ H_\text{p} $ \textit{ad hoc}.

On the other hand, the need to introduce the cutoff frequency $ \omega_c $, mentioned in the main text, is related to the fact that even if very large (as compared to the probe), the sample is finite and thus, it has a maximum energy. The non-equilibrium steady state of the central oscillator will unavoidably depend on the choice of $ \omega_c $ but, as long as $ \omega_c \gg \omega_0 $, this dependence should be weak and not change its qualitative features \cite{PhysRevA.86.012110}. In particular, note that $ \omega_R^2 \coloneqq \frac{2}{\pi}\int\nolimits_{0}^{\infty}\dif \omega\, \tfrac{J(\omega)}{\omega} = 2\gamma\omega_c $.

\section{From the Heisenberg equations to the QLE}
\label{app:HtoL}

We can write down the Heisenberg equations of motion $\big( \frac{d}{dt}A(t)=i[H,A(t)]+\partial_t A(t) \big)$ for all degrees of freedom $ \{x,p,x_\mu, p_\mu\} $ of the total system $ H = H_\text{p} + H_\text{s} + H_\text{p--s} $. These read
\begin{subequations}
	\begin{align}
		\dot{x}&=p \label{x_dot}\\
		\dot{p}&=-\big(\omega_0^2+\omega_R^2\big)x-\sum\nolimits_{\mu}g_\mu x_\mu \label{p_dot}\\
		\dot{x_\mu}&=\frac{p_\mu}{m_\mu} \label{xmu_dot}\\
		\dot{p_\mu}&=-m_\mu\omega_\mu^2 x_\mu-g_\mu x \label{pmu_dot}.
	\end{align}
\end{subequations}

Differentiating Eq.~\eqref{xmu_dot} and inserting in it Eq.~\eqref{pmu_dot} yields $ \ddot{x}_\mu +\omega_\mu^2 x_\mu= -\frac{g_\mu}{m_\mu} x $, which results in
\begin{align}
	x_\mu(t) = x_\mu(t_0)\cos\omega_\mu (t-t_0) + \frac{p_\mu(t_0)}{m_\mu\omega_\mu}\sin\omega_\mu (t-t_0) \nonumber\\
	- \frac{g_\mu}{m_\mu\omega_\mu}\int\nolimits_{t_0}^{t} \dif s \sin\omega_\mu(t-s)\,x(s).
	\label{xmu_t}
\end{align}

Similarly, one can differentiate Eq.~\eqref{x_dot} and use Eqs.~\eqref{p_dot} and \eqref{xmu_t} to eliminate $ \dot{p} $ and $ x_\mu $. This results in the following integro-differential equation
\begin{align}
	\ddot{x}+\left(\omega_0^ 2+\omega_R^ 2\right)x - \int\nolimits_{t_0}^t ds \sum\nolimits_\mu\frac{g_\mu^2}{m_\mu\omega_\mu}\sin\omega_\mu(t-s)\, x(s)
	\nonumber\\=-\sum\nolimits_{\mu}g_\mu\left(x_\mu(t_0)\cos\omega_\mu(t-t_0)\right. \nonumber\\
	+\left.\frac{p_\mu(t_0)}{m_\mu\omega_\mu}\sin\omega_\mu (t-t_0)\right).
	\label{qle_raw}
\end{align}

This is the quantum Langevin equation (QLE) for our probe. Since we are interested in the steady state of the central oscillator, we may let $ t_0\rightarrow-\infty $ without loss of generality. Defining the stochastic quantum force 
\begin{align}
	F(t) \coloneqq  -\sum\nolimits_{\mu}g_\mu\left(x_\mu(t_0)\cos\omega_\mu(t-t_0)\right. \nonumber\\
	\left. +\frac{p_\mu(t_0)}{m_\mu\omega_\mu}\sin\omega_\mu (t-t_0)\right),
	\label{quantum_force}
\end{align}
and the dissipation kernel 
\begin{align}
	\chi(t) &\coloneqq \sum\nolimits_\mu\frac{g_\mu^2}{m_\mu\omega_\mu}\sin\omega_\mu t\,\Theta(t)\nonumber\\
	&=\frac{2}{\pi}\int\nolimits_{0}^{\infty}\dif\omega~J(\omega)\sin\omega t~\Theta(t),
	\label{dissipation_kernel}
\end{align}
one may rewrite the QLE as
\begin{equation}
	\ddot{x}(t) + \left(\omega_0^2+\omega_R^2 \right) x(t) - x(t) \ast \chi(t) = F(t),
	\label{qle}
\end{equation}
where $\ast$ denotes convolution. Note that, so far, the initial state of the sample has not been specified and is thus completely general. In Sec.~\ref{app:FDR} below we shall adopt a thermal equilibrium preparation.

\section{Steady-state solution of the QLE}
\label{app:solvingQLE}

As explained in the main text, any Gaussian state (such as the steady state of the probe) is fully characterized by its first and second-order moments. In the case of a single-mode Gaussian state, these latter can be arranged in the $ 2 \times 2 $ real and symmetric covariance matrix $ \boldsymbol{\sigma} $. We therefore must be able to compute objects like $ \langle \{x(t'),x(t'')\}\rangle $, $ \langle \{p(t'),p(t'')\} \rangle $ and $ \langle\{ x(t'),p(t'')\} \rangle $ from Eq.~\eqref{qle}. Let us start by taking its Fourier transform, which gives
\begin{align}
	&-\omega^2\tilde{x} + (\omega_0^2+\omega_R^2)\tilde{x} + \tilde{x}\,\tilde{\chi} = \tilde{F}
	\nonumber\\
	 &\Rightarrow \tilde{x}(\omega) = \frac{\tilde{F}(\omega)}{\omega_0^2+\omega_R^2-\omega^2-\tilde{\chi}(\omega)} \coloneqq \alpha(\omega)^{-1}\tilde{F}(\omega).
	\label{qle_fourier}
\end{align}

Note that 
\begin{align}
	&\frac{1}{2}\langle \{x(t'),x(t'')\} \rangle  \nonumber\\ 
	&=\frac{1}{2}\int\limits_{-\infty}^{\infty} \frac{\dif\omega'}{2\pi}\,e^{-i\omega' t'}\int\limits_{-\infty}^{\infty}\frac{\dif\omega''}{2\pi}\,e^{-i\omega'' t''} \langle\{ \tilde{x}(\omega'),\tilde{x}(\omega'') \}\rangle\nonumber\\
	&=\frac{1}{2}\int\limits_{-\infty}^{\infty}\frac{\dif\omega'}{2\pi}\,e^{-i\omega' t'}\int\limits_{-\infty}^{\infty}\frac{\dif\omega''}{2\pi}e^{-i\omega'' t''}\,\alpha(\omega')^{-1}\alpha(\omega'')^{-1} \nonumber\\ &\hspace{40mm}\times\langle\{\tilde{F}(\omega'),\tilde{F}(\omega'')\} \rangle_T .
	\label{covariances_raw}
\end{align}
Therefore, all what is left is to find the analytical expression of the power spectrum of the sample $ 2^{-1}\langle \{\tilde{F}(\omega'),\tilde{F}(\omega'')\} \rangle_T $ and of the Fourier transform of the susceptibility $ \tilde{\chi}(\omega) $, which appears in $ \alpha(\omega) $. Note that the Fourier transform of all first order moments will be proportional to $ \langle\tilde{F}(\omega) \rangle_T  $ which is identically zero [cf. Eq.~\eqref{quantum_force}]. Hence, the steady states of the central oscillator will be \textit{undisplaced} Gaussians. With the subscript in $ \langle\cdots\rangle_T $, we emphasize that the average is taken over the initial Gibbs state of the sample.  

\subsection{The fluctuation-dissipation relation}
\label{app:FDR}

Let us start by computing $ \frac12\langle\{\tilde{F}(\omega'),\tilde{F}(\omega'')\}\rangle_T = \text{Re}\,{\langle \tilde{F}(\omega')\tilde{F}(\omega'') \rangle_T}$ from Eq.~\eqref{quantum_force}. Taking into account that $ \langle x_\mu(t_0)x_\mu'(t_0) \rangle_T = \delta_{\mu\mu'} (2 m_\mu\omega_\mu)^{-1}[1+2n_\mu(T)]$, $ \langle p_\mu(t_0)p_\mu'(t_0) \rangle_T = \delta_{\mu\mu'} \frac12 m_\mu\omega_\mu [1+2n_\mu(T)]$ and $ \langle x_\mu(t_0) p_\mu(t_0) \rangle_T = \langle p_\mu(t_0) x_\mu(t_0) \rangle_T^* = i/2$, one has
\begin{align}
	\frac{1}{2}\langle\{\tilde{F}(t'),\tilde{F}(t'')\}\rangle_T	&=\frac{1}{\pi}\sum_{\mu}\frac{\pi g_\mu^2}{2 m_\mu \omega_\mu}[1+2 n_\mu(T)]\nonumber\\
	&\times\,\bigg[\cos{\omega_\mu(t'-t_0)}\cos{\omega_\mu(t''-t_0)}\nonumber\\
	&\hspace{10mm}+\sin{\omega_\mu(t'-t_0)}\sin{\omega_\mu(t''-t_0)}\bigg] \nonumber\\
	&= \frac{1}{\pi}\int\limits_{0}^{\infty}\dif\omega\, J(\omega)\coth{\frac{\omega}{2T}}\cos{\omega(t'-t'')}
	\label{powerspectrum_raw},
\end{align}
where we have used $ 2n_\mu(T) + 1 = \coth{(\omega_\mu/2T)} $, which follows from the definition of the bosonic thermal occupation number $ n_\mu(T) \coloneqq [\exp{(\omega/2T)}-1]^{-1}  $. Now, taking the Fourier transform of Eq.~\eqref{powerspectrum_raw} yields
\begin{widetext}
\begin{align}
	\frac{1}{2}\langle\{\tilde{F}(\omega'),\tilde{F}(\omega'')\}\rangle_T &= 2\pi\int\limits_{-\infty}^{\infty}\frac{\dif t'}{2\pi}\,e^{i\omega' t'}\int\limits_{-\infty}^{\infty}\frac{\dif t''}{2\pi}\,e^{i\omega'' t''}\int\limits_{0}^{\infty}\dif\omega\,J(\omega)\coth{\frac{\omega}{2T}}\left(e^{i\omega(t'-t'')}+e^{-i\omega(t'-t'')}\right) \nonumber\\
	&= 2\pi\int\limits_{-\infty}^{\infty}\frac{\dif t'}{2\pi}\int\limits_{-\infty}^{\infty}\frac{\dif t''}{2\pi}\int\limits_{0}^{\infty}\dif\omega\,J(\omega)\coth{\frac{\omega}{2\pi}}\left(e^{it'(\omega+\omega')}e^{it''(\omega''-\omega)}+e^{it'(\omega'-\omega)}e^{it''(\omega''+\omega)}\right) \nonumber\\
	&= 2\pi\int\limits_{0}^{\infty}\dif\omega\,J(\omega)\coth{\frac{\omega}{2T}}\left[\delta(\omega+\omega')\delta(\omega''-\omega)+\delta(\omega'-\omega)\delta(\omega''+\omega)\right] \nonumber\\
	&= 2\pi\,\delta(\omega'+\omega'') \coth{\frac{\omega'}{2T}}\left[ J(\omega')\,\Theta(\omega') - J(-\omega')\,\Theta(-\omega') \right],
	\label{powerspectrum_raw2}
\end{align} 
\end{widetext}
where we have used the identity $ \int_{-\infty}^{\infty} \dif t\, e^{i\omega t} = 2\pi\,\delta(\omega)$. On the other hand, we may find $ \text{Im}\,\tilde{\chi}(\omega) $ from Eq.~\eqref{dissipation_kernel}. Note that
\begin{widetext}
\begin{align}
	\text{Im}\,\tilde{\chi}(\omega)&=\text{Im}\, \sum_\mu\frac{g_\mu^2}{m_\mu\omega_\mu}\int\limits_{-\infty}^{\infty}\dif t\,e^{i\omega t}\,\Theta(t)\,\sin{\omega_\mu t} =
	\sum_\mu\frac{g_\mu^2}{m_\mu\omega_\mu}\int\limits_{0}^{\infty}\dif t\,\sin{\omega t}\sin{\omega_\mu t} \nonumber\\
	&=-\frac{1}{4}\sum_\mu\frac{g_\mu^2}{m_\mu\omega_\mu}\int\limits_{0}^{\infty}\dif t\,[e^{i(\omega+\omega_\mu)t}-e^{i(\omega-\omega_\mu)t}-e^{i(-\omega+\omega_\mu)t}+e^{-i(\omega+\omega_\mu)t}] \nonumber\\
	&= -\frac{1}{4}\sum_\mu\frac{g_\mu^2}{m_\mu\omega_\mu}\left(\int\limits_{-\infty}^{\infty}\dif t\, e^{i(\omega+\omega_\mu)t} - \int\limits_{-\infty}^{\infty}\dif t\, e^{i(\omega-\omega_\mu)t} \right) \nonumber\\
	&= \frac{\pi}{2}\sum_\mu\frac{g_\mu^2}{m_\mu\omega_\mu}[\delta(\omega-\omega_\mu)-\delta(\omega+\omega_\mu)]=
	\int\limits_{0}^{\infty} \dif\omega'\, J(\omega') [\delta(\omega-\omega')-\delta(\omega+\omega')]\nonumber\\
	&=J(\omega)\,\Theta(\omega)-J(-\omega)\,\Theta(-\omega).
	\label{imchi}
\end{align} 
\end{widetext}
Hence the fluctuation-dissipation relation $ \langle\{\tilde{F}(\omega'),\tilde{F}(\omega'')\}\rangle = 4\pi\,\delta(\omega'+\omega'') \coth{(\omega'/2T)}\,\text{Im}\,\tilde{\chi}(\omega') $. When it comes to its real part, the calculation is not so straightforward. Recall from Eq.~\eqref{dissipation_kernel} that the response function $ \chi(t) $ is \textit{causal} due to the accompanying Heaviside step function. Causal response functions have analytic Fourier transform in the upper-half of the complex plane and therefore, the Kramers-Kronig relations hold \cite{weiss2008quantum}. In particular
\begin{equation}
	\text{Re}\,{\tilde{\chi}(\omega)} =\frac{1}{\pi}\,\text{P}\,\int\limits_{-\infty}^{\infty}d\omega'\,\frac{\text{Im}\,{\tilde{\chi}(\omega')}}{\omega'-\omega}\coloneqq\mathcal{H}\,\text{Im}\,{\tilde{\chi}(\omega)},
	\label{kramers_kronig}
\end{equation}   
where we have introduced the Hilbert transform $ g(y) = \mathcal{H}\,f(x) \coloneqq \pi^{-1}\,\text{P}\,\int_{-\infty}^{\infty}dx\,f(x)/(x-y) $ \cite{bateman1954tables2}, and $ \text{P} $ denotes Cauchy principal value.

\begin{widetext}
\section{Dissipation kernel for Ohmic and super-Ohmic spectral densities with exponential cutoff}
\label{app:kernel}

We will now obtain $ \re{\tilde{\chi}(\omega)} $ for two instances of the family of spectral densities $ J_s(\omega)\coloneqq \frac{\pi}{2}\gamma \omega^s\omega_c^{1-s} e^{-\omega/\omega_c} $, namely $ s=1 $ (Ohmic case) and $ s=2 $ (super-Ohmic case) \cite{PhysRevE.91.062123}. To begin with, let us list four useful properties of the Hilbert transform that we shall use in what follows
\begin{subequations}
	\begin{align}
	&	f(-a x) \xmapsto{\mathcal{H}} -g(-a y)   a>0 \label{hilbert_properties1}\\
	&	x f(x)  \xmapsto{\mathcal{H}} y\,g(y) + \frac{1}{\pi}\int\limits_{-\infty}^{\infty} dx\,f(x) \label{hilbert_properties2}\\
	&	\exp{(-a\vert x \vert)} \xmapsto{\mathcal{H}} \frac{1}{\pi} \text{sign}\,y\,\left[e^{a\vert y \vert}\,\text{Ei}(-a\vert y \vert) - e^{-a\vert y \vert}\,\overline{\text{Ei}}(a\vert y \vert)\right], \hspace{3mm}  a>0 \label{hilbert_properties3} \\
	&	\text{sign}\,x\,\exp{(-a\vert x \vert)} \xmapsto{\mathcal{H}} -\frac{1}{\pi} \,\left[\exp{(a\vert y \vert)}\,\text{Ei}(-a\vert y \vert) + \exp{(-a\vert y \vert)}\,\overline{\text{Ei}}(a\vert y \vert)\right], \hspace{2mm} a>0 \label{hilbert_properties4},
	\end{align}  
	\label{hilbert_properties}
\end{subequations}
where $ \text{Ei}(x) \coloneqq -\int_{-x}^{\infty}dt\, t^{-1}e^{-t}$ is the exponential integral, and $ \overline{\text{Ei}}(x) $ denotes its principal value.

\subsubsection{Ohmic case ($s=1$)}

According to Eqs.~\eqref{kramers_kronig} and \eqref{imchi}, one has
\begin{equation}
	\re{\tilde{\chi}(\omega)}=\frac{\pi\gamma}{2}\big\{\mathcal{H}[\Theta(\omega')\,\omega' \exp{(-\omega'/\omega_c)}](\omega) - \mathcal{H}[-\Theta(-\omega')\,\omega' \exp{(\omega'/\omega_c)}](\omega)\big\}.
	\label{rechi_step1}
\end{equation}

Using Eqs.~\eqref{hilbert_properties1} and \eqref{hilbert_properties2}, this rewrites as
\begin{multline}
	\re{\tilde{\chi}(\omega)}=\frac{\pi\gamma}{2}\big\{\mathcal{H}[\Theta(\omega')\,\omega' \exp{(-\omega'/\omega_c)}](\omega) + \mathcal{H}[\Theta(\omega')\,\omega' \exp{(-\omega'/\omega_c)}](-\omega)\big\} \\
	= \frac{\pi\gamma}{2}\big\{\omega\,\mathcal{H}[\Theta(\omega')\, \exp{(-\omega'/\omega_c)}](\omega) - \omega\,\mathcal{H}[\Theta(\omega')\,\exp{(-\omega'/\omega_c)}](-\omega) + \frac{2\omega_c}{\pi}\big\}.
	\label{rechi_step2}
\end{multline} 
Now, using first Eq.~\eqref{hilbert_properties1} again, and then Eq.~\eqref{hilbert_properties3}, one finds
\begin{equation}
	\re{\tilde{\chi}(\omega)} = \gamma\omega_c + \frac{\pi\gamma}{2}\omega\,\mathcal{H}[\exp{(-\vert\omega'\vert/\omega_c)}](\omega)
	=\gamma\omega_c - \frac{\gamma}{2}\omega\,\big[\exp{(-\omega/\omega_c)}\,\overline{\text{Ei}}(\omega/\omega_c)-\exp{(\omega/\omega_c)\,\text{Ei}(-\omega/\omega_c)}\big],
	\label{rechi}
\end{equation}
which can also be expressed in terms of the incomplete Euler's Gamma function $ \Gamma(0,x) = -\text{Ei}(-x) $.

\subsubsection{Super-Ohmic case ($s=2$)}\label{app:super-ohmic}

Using the properties of Eq.~\eqref{hilbert_properties} it is also straightforward to obtain $ \re{\tilde{\chi}(\omega)} $ in the case of $ s = 2 $:
\begin{align}
	\begin{split}
		\re{\tilde{\chi}(\omega)} &= \frac{\pi\gamma}{2\omega_c}\big\{\mathcal{H}[\Theta(\omega')\,\omega'^2\exp{(-\omega'/\omega_c)}](\omega) - \mathcal{H}[\Theta(-\omega')\,\omega'^2\exp{(\omega'/\omega_c)}](\omega) \big\} \\
		&=\frac{\pi\gamma}{2\omega_c}\big\{\mathcal{H}[\Theta(\omega')\,\omega'^2\exp{(-\omega'/\omega_c)}](\omega) + \mathcal{H}[\Theta(\omega')\,\omega'^2\exp{(-\omega'/\omega_c)}](-\omega) \big\} \\
		&=\frac{\pi\gamma}{2\omega_c}\big\{\omega\,\mathcal{H}[\Theta(\omega')\,\omega'\exp{(-\omega'/\omega_c)}](\omega) -  \omega\,\mathcal{H}[\Theta(\omega')\,\omega'\exp{(-\omega'/\omega_c)}](-\omega) + \frac{2\omega_c^2}{\pi}\big\} \\
		&= \gamma\omega_c + \frac{\pi\gamma}{2\omega_c}\big\{\omega^2\,\mathcal{H}[\Theta(\omega')\,\exp{(-\omega'/\omega_c)}](\omega) +  \omega^2\,\mathcal{H}[\Theta(\omega')\,\exp{(-\omega'/\omega_c)}](-\omega) \big\} \\		
		&= \gamma\omega_c + \frac{\pi\gamma}{2\omega_c}\omega^2\,\mathcal{H}[\text{sign}\,\omega\,\exp{(-\vert\omega'\vert/\omega_c)}](\omega) \\
		&= \gamma\omega_c - \frac{\gamma}{2\omega_c}\omega^2\,\big[ \exp{(-\omega/\omega_c)}\,\overline{\text{Ei}}(\omega/\omega_c) + \exp{(\omega/\omega_c)\,\text{Ei}(-\omega/\omega_c)} \big].
		\label{rechi_s2}
	\end{split}
\end{align}

\section{Calculation of the steady-state covariances}
\label{app:covariance}

Now we have all the ingredients to compute the steady-state covariances of the central oscillator. Note that 
\begin{align}
	\frac{1}{2}\langle\{x(t'),x(t'')\}\rangle &=\frac{1}{2}\int\limits_{-\infty}^{\infty}\frac{\dif\omega'}{2\pi}\,e^{-i\omega' t'}\int\limits_{-\infty}^{\infty}\frac{\dif\omega''}{2\pi}\,e^{-i\omega'' t''} \alpha(\omega')^{-1}\alpha(\omega'')^{-1}\langle\{\tilde{F}(\omega'),\tilde{F}(\omega'')\}\rangle_T \\
	&=\int\limits_{-\infty}^{\infty}\frac{\dif\omega'}{2\pi}\,e^{-i\omega' t'}\int\limits_{-\infty}^{\infty} \dif\omega''\,e^{-i\omega'' t''} \alpha(\omega')^{-1}\alpha(\omega'')^{-1} [J(\omega')\Theta(\omega')-J(-\omega')\Theta(-\omega')]\coth{\frac{\omega'}{2T}}\,\delta(\omega'+\omega'') \\
	&= \int\limits_{-\infty}^{\infty}\frac{\dif\omega'}{2\pi}\,e^{-i\omega' (t'-t'')} \alpha(\omega')^{-1}\alpha(-\omega')^{-1} [J(\omega')\Theta(\omega')-J(-\omega')\Theta(-\omega')]\coth{\frac{\omega'}{2T}}.
	\label{eq:xx}
\end{align}
This gives a closed expression for the position-position covariance. Note that, since $ \tilde{p}(\omega) = -i\omega\,\tilde{x}(\omega)$, one has $2^{-1}\langle \{\tilde{p}(\omega'),\tilde{x}(\omega'') \}\rangle = 0 $ and 
\begin{equation}
	\frac{1}{2}\langle \{ p(t'), p(t'') \} \rangle = \int\limits_{-\infty}^{\infty}\frac{\dif\omega'}{2\pi}\,e^{-i\omega' (t'-t'')}\,\omega'^2\, \alpha(\omega')^{-1}\alpha(-\omega')^{-1} [J(\omega')\Theta(\omega')-J(-\omega')\Theta(-\omega')]\coth{\frac{\omega'}{2T}}.
	\label{eq:pp}
\end{equation} 
\end{widetext}
Therefore, we have fully characterized the steady state of a single harmonic oscillator in a bosonic bath. Note that the \textit{only} underlying assumption is that the sample was prepared in an equilibrium state at temperature $ T $. Specifically, this was required when evaluating the correlators $ \langle \{x_\mu(t_0),x_\mu(t_0) \} \rangle_T $ and $ \langle \{p_\mu(t_0),p_\mu(t_0) \} \rangle_T $ in Eq.~\eqref{powerspectrum_raw}. Otherwise, our calculation is \textit{completely general}. For a non-equilibrium sample, one would only need to recalculate Eqs.~\eqref{powerspectrum_raw} and \eqref{powerspectrum_raw2}. 

\subsection{Explicit calculation for Ohmic spectral density with Lorentz-Drude cutoff}
\label{app:explicit-ohmic}

The integrals in Eqs.~\eqref{eq:xx} and \eqref{eq:pp} are easy to evaluate numerically. However, when dealing with the simple Ohmic spectral density with Lorentz-Drude cutoff introduced in the main text as $ J(\omega) = 2\gamma\omega_c^2\omega/(\omega^2+\omega_c^2) $, it is possible to calculate the covariances analytically. This will allow us to get some insight into the temperature-dependence of the covariances at very low $ T $ and about the squeezing in the position quadrature described in the main text. 

Let us start calculating $ \langle x^2 \rangle $. For our choice of spectral density Eq.~\eqref{eq:xx} reads
\begin{widetext}
\begin{equation}
	\langle x^2 \rangle = \frac{\gamma\omega_c^2}{\pi}\int_{\infty}^{\infty} \dif\omega\frac{\frac{\omega}{\omega^2+\omega_c^2}\coth{\frac{\omega}{2T}}}{(\omega_0^2-\omega^2+2\gamma\omega_c-\frac{2\gamma\omega_c^2}{\omega_c-i\omega})(\omega_0^2-\omega^2+2\gamma\omega_c-\frac{2\gamma\omega_c^2}{\omega_c+i\omega})},
	\label{eq:integrand}
\end{equation}
which can be re-written as 
\begin{equation}
	\langle x^2 \rangle = \frac{2T\gamma\omega_c^2}{\pi}\left(\sum_{n=1}^\infty\int_{-\infty}^{\infty}\dif\omega\frac{\omega^2}{h_4(\omega)h_4(-\omega)} + \int_{-\infty}^{\infty} \frac{\dif \omega}{h_3(\omega)h_3(-\omega)} \right),
	\label{eq:integrand_2}
\end{equation}
\end{widetext}
where $ h_4(\omega) \coloneqq (\omega-i\nu_n)[(\omega_0^2-\omega^2+2\gamma\omega_c)(\omega_c+i\omega)-2\gamma\omega_c^2] $, $ h_3(\omega) = (\omega_0^2-\omega^2+2\gamma\omega_c)(\omega_c+i\omega)-2\gamma\omega_c^2 $, and owing to the identity $ \coth{\frac{\omega}{2T}} = 2\sum_{n=1}^\infty \frac{2T\omega}{\nu_n^2 + \omega^2} + \frac{2T}{\omega} $, where $ \nu_n\coloneqq 2\pi T n $ are the Matsubara frequencies.

Integrals such as those in Eq.~\eqref{eq:integrand_2} can be evaluated using the following formula \cite{gradshteyn1980ryzhik}
\begin{equation}
	\int_{-\infty}^{\infty}\dif x\frac{g_n(x)}{h_n(x)h_n(-x)} = \frac{i \pi}{a_0}\frac{\det{\mathsf{M}_n}}{\det{\mathsf{\Delta}_n}},
	\label{eq:formula}
\end{equation}
where $ g_n(x) \coloneqq b_0 x^{2n-2} + b_1 x^{2n-4} + \cdots + b_{n-1} $ and $ h_n(x) \coloneqq a_0 x^n + a_1 x^{n-1} + \cdots + a_n $ and the matrices $ \mathsf{\Delta}_n $ and $ \mathsf{M}_n $ are defined as
\begin{equation}
	\mathsf{\Delta}_n\coloneqq\begin{pmatrix}
		a_1 & a_3 & \cdots & 0  \\
		a_0 & a_2 & \cdots  & 0 \\
		0 & a_1 & \cdots & 0 \\
		\vdots  & \vdots & \ddots &  \vdots \\
		0 & 0 & \cdots  & a_n 
	\end{pmatrix},
	\qquad
	\mathsf{M}_n\coloneqq\begin{pmatrix}
		b_0 & b_1 & \cdots & b_{n-1}  \\
		a_0 & a_2 & \cdots  & 0 \\
		0 & a_1 & \cdots & 0 \\
		\vdots  & \vdots & \ddots &  \vdots \\
		0 & 0 & \cdots  & a_n 
	\end{pmatrix}.
	\label{eq:matrices_formula}
\end{equation}
For \eqref{eq:formula} to be valid, $ h_n(x) $ must have all its roots in the upper half of the complex plane, which is the case for us. The covariance $ \langle x^2 \rangle $ thus rewrites as
\begin{align}
	\langle x^2 \rangle = 2\sum_{n=1}^\infty\frac{T(\nu_n + \omega_c)}{\nu_n(\nu_n^2 + \omega_0^2) + (\nu_n^2 + 2\gamma\nu_n + \omega_0^2)\omega_c} + \frac{1}{2\omega_0^2}.
	\label{eq:sum_formal}
\end{align}

To proceed further, we shall resort to the \textit{digamma} function $ \psi(z) $, defined as the logarithmic derivate of Euler's gamma function \cite{abramowitz1970handbook}; that is $ \psi(z)\coloneqq\frac{\dif}{\dif z}\ln{\Gamma(z)} $, where $ \Gamma(z)\coloneqq\int_0^\infty\dif t\,t^{z-1}e^{-t} $. The digamma function satisfies the following identity \cite{abramowitz1970handbook}
\begin{align}
	\sum_{n=1}^\infty \frac{G(n)}{H(n)} = \sum_{n=0}^\infty \frac{G(n+1)}{H(n+1)} &= \sum_{n=0}^\infty\sum_{m=1}^N\frac{c_m}{n-d_m} \nonumber\\
	&= -\sum_{m=1}^N c_m \psi(-d_m),
	\label{eq:digamma_sum_identity}
\end{align}
where $ G(n) $ and $ H(n) $ are polynomials in $ n $, $ d_m $ are the $ N $ roots (assumed to be simple) of $ H(n+1) $, and $ c_m $ are the coefficients of the simple-fraction decomposition of $ G(n+1)/H(n+1) $ ($ \sum_{m=1}^N c_m = 0 $). In our specific case, the $ c_m $ evaluate to
\begin{align}
	&\hspace{-1mm}c_m =\nonumber\\ 
	&\frac{1}{2\pi}\frac{\nu_1(d_m + 1) + \omega_c}{\omega_0^2 + 2\gamma\omega_c + \nu_1 (d_m + 1) + \nu_1 (d_m + 1)[3\nu_1(d_m + 1)+2\omega_c]},
	\label{eq:coeff_simple_fraction}
\end{align}
and the $ d_m $ are the three solutions to
\begin{align}
	\nu_1^3(d+1)^3+\nu_1^2(d+1)^2\omega_c+\nu_1(d+1)(\omega_0^2+2\gamma\omega_c)\nonumber\\
	+\omega_0^2\omega_c = 0
	\label{eq:roots}.
\end{align}

Therefore, the covariance $ \langle x^2\rangle $ is
\begin{equation}
	\langle x^2 \rangle = \frac{1}{2\omega_0^2} - 2\sum_{m=1}^3 c_m \psi(-d_m).
	\label{eq:xx_analytical}
\end{equation}

Similarly, the momentum covariance can be found to be
\begin{equation}
	\langle p^2 \rangle = \frac12 - 2\sum_{m=1}^3 c'_m \psi_m(-d_m),
	\label{eq:pp_analytical}
\end{equation}
where the coefficients $ c'_m $ are now given by
\begin{equation}
	c'_m = \frac{\nu_1}{\pi}\frac{\omega_0^2\omega_c+\nu_1(d_m+1)(\omega_0^2+2\gamma\omega_c)}{\omega_0^2 + 2\gamma\omega_c+\nu_1(d_m+1)[3\nu_1(d_m+1)+2\omega_c]}.
	\label{eq:coeff_simple_fraction_pp}
\end{equation}

It must be noted that Eqs.~\eqref{eq:xx_analytical} and \eqref{eq:pp_analytical} are \textit{exact}, though not very informative. In the next section we will try to simplify their expressions by taking the low temperature limit.

\subsubsection{Low $ T $ and large $ \omega_c $ limit}
\label{app:explicit-ohmic-limit}

Let us consider again Eq.~\eqref{eq:roots}. To begin with, let us assume that $ \gamma/\omega_c \lll 1 $ so that $ d_m \simeq d_m^{(0)} + \frac{\gamma}{\omega_c} d_m^{(1)} $. One thus has
\begin{align}
	d_{1,2} &= -\left(1+\frac{\gamma\omega_c^2}{\nu_1(\omega_0^2+\omega_c^2)}\right)\nonumber\\
	&\pm i\,\frac{\omega_0}{\nu_1}\frac{\omega_0^2+\omega_c(\gamma+\omega_c)}{\omega_0^2+\omega_c^2} + \pazocal{O}\left(\frac{\gamma}{\omega_c}\right)^2 
	\nonumber\\
	d_3 &= -\left(1+\frac{\omega_c}{\nu_1}\right) + \frac{2\gamma\omega_c^2}{\nu_1(\omega_0^2+\omega_c^2)} + \pazocal{O}\left(\frac{\gamma}{\omega_c}\right)^2.
	\label{eq:roots_approx}
\end{align}

Notice that the roots diverge as $ T\rightarrow 0 $ due to the $ \nu_1 $ appearing in the denominators. It is thus possible to replace the digamma by the first term in its asymptotic expansion $ \psi(z)\sim\ln{z} $. 

Eqs.~\eqref{eq:xx_analytical} and \eqref{eq:pp_analytical} can be further simplified by retaining terms only up to first order in $ \omega_0/\omega_c $ and $ T/\omega_0 $. When expanding the expressions above, care must be taken with the divergence of terms proportional to $ \ln{\frac{\omega_c}{\omega_0}} $. One eventually arrives to the following approximate covariances
\begin{subequations}
	\begin{equation}
		\langle x^2 \rangle \simeq \frac{1}{2\omega_0} -\frac{1}{2\omega_0} \left(\frac{2\gamma}{\pi\omega_0}+\frac{2T}{\omega_0}+\frac{4\gamma\omega_0}{\pi\omega_c^2}\ln{\frac{\omega_c}{\omega_0}}\right)
		\label{eq:xx_approx}
	\end{equation}
	\begin{equation}
		\langle p^2 \rangle \simeq \frac{\omega_0}{2} + \frac{\omega_0}{2}\left[ \frac{4\gamma}{\pi\omega_0}\ln{\frac{\omega_c}{\omega_0}} + \frac{3\gamma}{\omega_c} - \left(\frac{2T}{\omega_0}+\frac{2\gamma}{\pi\omega_0}\right)\right].
		\label{eq:pp_approx}
	\end{equation}
	\label{eq:covariances_approx}
\end{subequations}

From Eq.~\eqref{eq:xx_approx} we can see how the variance in the position quadrature is reduced below its thermal equilibrium value of $ \langle x^2 \rangle_T = (2\omega_0)^{-1}\coth{\frac{\omega_0}{2T}}\sim (2\omega_0)^{-1} $, as noted in the main text. On the other hand, the `quantum correction' over $ \langle p^2 \rangle_T $ [i.e., the bracketed term in Eq.~\eqref{eq:pp_approx}] is dominated by the non-perturbative logarithmic divergence, and will therefore be positive. 	In particular, for strong dissipation, i.e. $ \gamma/\omega_0 \gtrsim 1 $, $ \langle p^2 \rangle \simeq \langle p^2 \rangle_T + \frac{2\gamma}{\pi}\ln{\frac{\omega_c}{\omega_0}} $ and hence $ \partial_T\langle p^2 \rangle \simeq 0 $. On the contrary, there is no reason to drop the temperature dependence of $ \langle x^2 \rangle $ in the strong dissipation regime. This intuitively justifies our observation that the dispersion in the position quadrature exhibits a quasi-optimal thermal sensitivity in the ultra-cold strongly-coupled regime, whilst the dispersion in momentum performs very poorly as a temperature estimator. 

Unfortunately, Eqs.~\eqref{eq:covariances_approx} are unsuitable to derive a qualitatively accurate and equally simple analytical expression for the low-temperature QFI. One should proceed instead directly from Eqs.~\eqref{eq:xx_analytical} and \eqref{eq:pp_analytical} and expand the resulting expression again to first order in the small parameters $ \gamma/\omega_c $, $ \omega_0/\omega_c $, and $ T/\omega_0 $. Although this is in principle straightforward, the algebra quickly becomes unmanageable.

\section{Dependence of the normal-mode frequencies on the coupling strength in a `star system'}
\label{app:freqsvscoupling}

Let us consider a finite \textit{star system} with $ N $ modes. As already explained in the main text, this will be comprised of a central harmonic oscillator of bare frequency $ \omega_0 $ (playing the role of the probe), dissipatively coupled to $ N-1 $ independent peripheral oscillators with arbitrary frequencies $ \omega_{\mu\in\{1,\cdots,N-1 \}} $ (representing the sample). We will choose linear probe-sample couplings of the form $ x\,G\,\sum_{\mu=1}^{N-1}g_\mu x_\mu $. Therefore, adjusting $ G $ simply amounts to rescaling the probe-sample interaction without changing the overall frequency distribution of the couplings. This is exactly what happens when the dissipation strength $ \gamma $ is tuned in the spectral density $ J(\omega) $ of the continuous Caldeira-Leggett model from the main text. Note that we also allow for an \textit{arbitrary} frequency-distribution of the coupling constants $ g_\mu $.

Hence, the total $ N $-particle Hamiltonian may be written as $ \hat H = \frac{1}{2} \bar{x}^{\mathsf{t}}\mathsf{V}\bar{x} + \frac12 \vert\bar{p}\vert^2 $. Here, the $ N $-dimensional vectors $ \bar{x} $ and $ \bar{p} $ are $ \bar{x} = ( x, x_1,\cdots, x_{N-1} ) $ and $ \bar{p} = ( p, p_1,\cdots, p_{N-1} ) $. For simplicity, we will take unit mass for all particles. The $ N \times N $ interaction matrix $ \mathsf{V} $ may thus be written as
\begin{equation}
	\mathsf{V} = 
	G\begin{pmatrix}
		G^{-1}\Omega_0^2 & g_1 & g_2 & \cdots  & g_{N-2} & g_{N-1} \\
		g_1 & G^{-1}\omega_1^2 & 0 & \cdots  & 0 & 0 \\
		g_2 & 0 & \omega^2_2 & \cdots  & 0 & 0 \\
		\vdots  & \vdots & \vdots & \ddots &  \vdots & \vdots \\
		g_{N-2} & 0 & 0 & \cdots  & G^{-1}\omega^2_{N-2} & 0 \\
		g_{N-1} & 0 & 0 & \cdots  & 0 & G^{-1}\omega^2_{N-1} \\ 
	\end{pmatrix}.
	\label{eq:interaction_matrix}
\end{equation}

The frequencies of the normal modes of the system are given by the square root of the $ N $ solutions $ \lambda_i $ of $ P_N(\lambda_i) = \vert\mathsf{V}-\lambda_i\mathbbm{1}\vert = 0 $. Note that we have shifted the frequency of the central oscillator $ \omega_0^2\rightarrow\Omega_0^2\coloneqq\omega_0^2 + \sum_{\mu} g_\mu^2/\omega_\mu^2 $ to ensure that all $ \lambda_i > 0 $.

While it is hard to obtain closed expressions for $ \lambda_i $, one may easily see the following: The frequencies of the modes above $ \Omega_0 $ increase with the coupling strength, whereas those of the modes below $ \Omega_0 $ decrease with $ G $ (i.e. $ \partial_G\lambda_i > 0 $ for $ \lambda_i > \Omega_0^2 $ and $ \partial_G\lambda_i < 0 $ for $ \lambda_i < \Omega_0^2 $). Indeed, expanding $ P_N(\lambda) $ by minors along the last row, yields the recurrence relation
\begin{equation}
	P_N(\lambda) = (\omega_{N-1}^2-\lambda) P_{N-1}(\lambda) - G^2\,g_{N-1}^2 \Pi_{k=1}^{N-2}(\omega_k^2-\lambda),
\end{equation}
which allows to rewrite the condition $ P_N(\lambda_i) = 0 $ as 
\begin{equation}
	\Omega_0^2-\lambda_i=\frac{1}{\prod_{l=1}^{N-1}\omega^2_l-\lambda_i}\sum_{k=1}^{N-1}G^2 g_k^2\prod_{l=1}^{N-1}\frac{\omega^2_l-\lambda_i}{\omega^2_k-\lambda_i}=\sum_{k=1}^{N-1}\frac{G^2 g_k^2}{\omega^2_k-\lambda_i}.
	\label{eq:eigenvalues_V}
\end{equation}

Consequently, the derivative of any eigenvalue $ \lambda_i $ with respect to the coupling strength $ G $ evaluates to
\begin{equation}
	\partial_G\lambda_i=-\frac{2G\sum_{k=1}^{N-1}g_k^2(\omega^2_k-\lambda_i)^{-1}}{1+\sum_{k=1}^{N-1}G^2 g_k^2(\omega^2_k-\lambda_i)^{-2}}.
	\label{eq:derivative_eigenvalues_V}
\end{equation}

Comparing Eqs.~\eqref{eq:eigenvalues_V} and \eqref{eq:derivative_eigenvalues_V} we can see that $ \partial_G\lambda_i > 0 $ for $ \lambda_i > \Omega_0^2 $, and that, on the contrary, $ \partial_G\lambda_i < 0 $ for $ \lambda_i < \Omega_0^2 $. 

Now consider the situation in which the star system is prepared in a Gibbs state at temperature $ T $. Paraphrasing the line of reasoning of the main text, if $ T $ happens to be so low that not even the fundamental mode is significantly populated, the thermal sensitivity of the entire system, and also that of the central temperature probe, vanishes. However, if we were to increase the coupling strength $ G $, the frequencies of the lowest normal modes would decrease monotonically. As a result, the first few modes could get thermally populated thus enabling temperature sensing. 

This intuition can be made more precise by explicitly writing the total QFI of the star system $ \pazocal{F}_T^{(\text{star})} $. Its global thermal state can be expressed as $ \rho_T \propto \exp{(-H/T)} = \bigotimes_{i=1}^N \varrho_T^{(i)} $, where $ \varrho_T^{(i)} $ stands for the Gibbs state of the normal mode at frequency $ \sqrt{\lambda_i} $. Since the QFI is additive with respect to tensor products, one has $ \pazocal{F}_T^{(\text{star})} = \sum_{i=1}^N\pazocal{F}_T^{(\text{eq})}\big(\sqrt{\lambda_i}\big) $, where the QFI for temperature estimation in a thermal mode $ \pazocal{F}_T^{(\text{eq})} $ was defined in the main text.

If the temperature $ T $ is low enough, only the terms corresponding to the lowest-frequency normal modes will contribute significantly to the sum in $ \pazocal{F}_T^{(\text{star})} $. Crucially, $ \pazocal{F}_T^{(\text{eq})}(\omega) $ also increases monotonically as $ \omega\rightarrow 0 $ which, in turn, entails a \textit{monotonic} increase of $ \pazocal{F}_T^{(\text{star})} $ with $ G $ at low $ T $. If, on the contrary, the temperature were large enough to thermally populate modes above $ \Omega_0 $, the situation would become less clear: The global QFI could either increase or decrease with $ G $. Due to its central position, the QFI of the reduced state of the probe qualitatively follows $ \pazocal{F}_T^{(\text{star})} $ (although $ \pazocal{F}_T \ll \pazocal{F}_T^{(\text{star})} $).


%

\end{document}